# Cepheid radii and the CORS method revisited

V. Ripepi[1], F. Barone[2], L. Milano[2], and G. Russo[1,3]

[1] Dipartimento di Fisica, Università della Calabria
I-87036 Arcavacata di Rende (CS) Italy.
[2] Dipartimento di Scienze Fisiche, Università di Napoli Federico II
pad. 20 Mostra d'Oltremare, I-80125 Napoli, Italy.
[3] INFN, Gruppo collegato di Cosenza, Dipartimento di Fisica
I-87036 Arcavacata di Rende (CS) Italy I-87036.
Affiliated to Osservatorio Astronomico di Capodimonte, Napoli.

December 22, 1995

**Abstract.** We have refined the CORS method, introduced in 1980 for the computation of the cepheid radii, in order to extend its applicability to recent and extensive sets of observations. The refinement is based on the computation, from observational data only, of one of the terms of the solving equation, previously based only on precise calibrations of photometric colours. A limited number of assumptions, generally accepted in the literature, is used.

New radii are computed for about 70 cepheids, and the resulting P-R relation is discussed.

## 1. Introduction

The importance of a proper knowledge of the radii of Cepheid variables is well known, in particular, but not only, in connection with the problem of the cosmic distance scale. Notwithstanding extensive studies by several authors (Caccin et al. 1981; Fernie 1984; Gieren 1986; Moffett & Barnes 1987; Gieren et al. 1989, Laney & Stobie 1995), there are still doubts on the correct period-radius relation. Almost all methods used for determining the radius of a cepheid from the photometric and spectroscopic (radial velocities) observations, are based on the classical Baade-Wesselink method (Wesselink, 1946). Gautschy (1987) pubblished a review of the different methods; this comparative study outlines that the CORS method (Caccin et al., 1981, Sollazzo et al., 1981) had the most solid physical basis, together with the surface brightness technique by Barnes & Evans (1976). However, the CORS method suffered from two major drawbacks that have so far

*Send offprint requests to*: G. Russo

prevented its application to most observational data; that is the mathematically difficult formulation, and the need of a good and complete calibration of the colours, able to derive detailed temperatures and gravity curves. In particular, the second problem is solved so far only for the VBLUW observations by Pel (1976,1978). The first problem is instead not a true problem, because, as we will show, the whole CORS method reduces itself to the fitting of the data and to the solution of an implicit equation which, once written in computer form, requires a fraction of a second to be solved in terms of the cepheid radius.

The purpose of the present paper is to introduce a modification of the CORS method, by taking into account the surface brightness method, in order to allow the determination of cepheid radii for a wider set of data than it was originally conceived.

In the following we will first recall the original surface brightness and CORS methods, and then we introduce its mofied version, discussing its characteristics. Last sections are devoted to the application of the method and to the discussion of the results.

## 2. The new method

### 2.1. The surface brightness method

The surface brightness method (Barnes & Evans, 1976), as used by Gieren et al. (1989), is based on the visual surface brightness $S_v$, defined as the following equivalent relations:

$$S_V = V + 5 \cdot \log \alpha \qquad (1)$$

$$S_V = -10 \cdot \log T_e - \text{B.C.} + \text{const.} \qquad (2)$$

tric corrections. Or, equivalently, the surface brightness parameter $F_v$, given by

$$F_V = \text{const.} - 0.1 \cdot S_V = \log T_e + 0.1 \cdot B.C. \tag{3}$$

An empirical relation has been found by Barnes & Evans (1976), which correlates $F_v$ to the Johnson $(V-R)_0$ index, corrected for interstellar reddening:

$$F_V = b + m \cdot (V-R)_0 \tag{4}$$

where the angular coefficient m is in turn a function of the pulsational period (Moffett & Barnes, 1987):

$$m = -0.370 + 0.004 \cdot \log P \tag{5}$$

The practical application starts from the (V-R) observations, which from (4) give $F_v$, which in turn from (3) give $S_v$, which in turn from (1) give $\alpha$. Expressing the linear diameter $D = 10^{-3} \cdot r \cdot \alpha$ (r being the distance in parsec) as $\Delta D + D_m$ ($\Delta D$ being the instantaneous displacement) and performing a regression analysis of $\alpha$ against $\Delta D$, one obtains the mean diameter $D_m$, as well as the distance r.

### 2.2. The original CORS Method

The CORS method also starts from (1), but proceeds mathematically by differentiating it with respect to the phase, multiplying by the colour index (B-V) and integrating over the whole cycle; after substituting:

$$\dot{R}(\phi) = -p \cdot P \cdot u(\phi) \tag{6}$$

it yelds to the following equation, in which $\phi$ is the phase

$$a \int_0^1 \log\{R_0(\phi) - p \cdot P \int_{\phi_0}^{\phi} u(\phi') \cdot d\phi'\}(B \dot{-} V)(\phi) \cdot d\phi + \\ -B + \Delta B = 0 \tag{7}$$

where $a = 5/\log_e 10$,

$$B = \int_0^1 (B-V)(\phi) \cdot \dot{V}(\phi) \cdot d(\phi) \tag{8}$$

$$\Delta B = \int_0^1 (B-V)(\phi) \cdot \dot{S_V}(\phi) \cdot d(\phi) \tag{9}$$

P is the period, u the radial velocities and p is a conversion factor (Parsons, 1972; Gieren et al. 1989). The practical application of the CORS method starts from a fitting of data with respect to the phase $\phi$ by means, e.g., of Fourier series. The data are given by the V magnitude, the (B-V) and (V-R) colour indexes, and the radial velocities u.

The fit is easily obtained with an interactive procedure on a computer terminal with graphical capabilities; afterwards, the fitted curves are used to compute in an automated way the term B, the term $\Delta$ B, the derivatives and eventually to solve eq. (7) to obtain $R_0$, the radius at an arbitrary phase $\phi_0$; $\phi_0$ is usually taken at the minimum of the radial velocity curve, but its choice is inessential; the mean radius comes from integrating twice eq. (6).

We will now show how the two methods can be used together, to obtain what we will call the modified CORS method.

From eq. (4) we can obtain $F_V$, and from it, we can obtain $S_V$, using also eq. (3) and (5):

$$\begin{aligned} S_V &= const. - 10 \cdot F_V \\ &= const. - 10 \cdot m \cdot (V-R)_0 - 10 \cdot b \\ &= const. + (3.7 - 0.04 \cdot \log P)(V-R)_0 \end{aligned} \tag{10}$$

The value of the constant is of no importance in our case, since we are interested in computing $\Delta B$; this term, given by eq. (9), is the area of the loop described by the star in the plane $S_V$ vs (B-V). If we make the transformation of variable from $S_V$ to $(V-R)_0$, given by eq. (10), then:

$$\begin{aligned} \Delta B &= \int_0^1 (B-V)(\phi) \cdot \dot{S_V}(\phi) \cdot d(\phi) \\ &= (3.7 - 0.04 \cdot \log P) \cdot \\ &\quad \cdot \int_0^1 (B-V)(\phi) \cdot (V \dot{-} R)_0(\phi) \cdot d(\phi) \end{aligned} \tag{11}$$

Since the derivative of $(V-R)_0$ with respect to the phase is, in first approximation, the same as the one of (V-R), we get from eq. (11), with $c = 3.7 - 0.04 \cdot \log P$:

$$\Delta B = c \int_0^1 (B-V)(\phi) \cdot (V \dot{-} R)(\phi) \cdot d(\phi) \tag{12}$$

Let us pose $C = \int_0^1 (B-V)(\phi) \cdot (V \dot{-} R)(\phi) \cdot d(\phi)$, wich represents the area of the loop described by the star in the plane (V-R) vs (B-V), we get to the final formulation, given the following observations:

*i)* the light curve
*ii)* the colour curve (B-V)
*iii)* the colour curve (V-R)
*iv)* the radial velocity curve

and the quantities, all based on the above observations:

*i)* B=area of the loop (B-V) vs V
*ii)* C=area of the loop (V-R) vs (B-V)

The radius $R_0$ at an arbitrary phase is obtained from the equation:

$$a \int_0^1 \log\{R_0(\phi) - p \cdot P \int_{\phi_0}^{\phi} u(\phi') \cdot d\phi'\}(B \dot{-} V)(\phi) \cdot d\phi + \\ -B + c \cdot C = 0 \tag{13}$$

where $c = 3.7 - 0.04 \cdot \log P$ and $p = 1.39 - 0.030 \cdot \log P$, if we follow Gieren et al. (1989), or $c=3.7$ and $p=1.36$ in our approximation.

radial velocity curve, since the first integration gives the radius curve and the second integration its mean value. Eq. (13) has to be solved by numerical methods, but this is easily accomplished with any computer, which nowdays is a common tool for any astronomer.

The assumptions and approximations which are behind this formulation are the following:

i) the observational data give a good coverage of the pulsational period, and do not show big noise, so that the Fourier fitting (or any other fitting which guarantees the periodicity of the data) is a good approximation to the observations.

ii) the photometric and spectroscopic observations are simultaneous, or at least separated by few tens of pulsational cycles, so that no phase-shift is present in the terms of eq. (13)

iii) the correlation of eq. (5), given by Moffett & Barnes (1987) represents a good approximation to the data.

iv) the proportionality between $\Delta B$ and C is valid; this was already proved in general terms by Onnembo et al., (1985), and therefore what we found here is just a confirmation, particularly valid for the colours (B-V) and (V-R).

v) the colours are not affected by other contributions, e.g. the presence of companions to the cepheid (see Russo et al. 1981).

## 3. Stability of the method

We want to study here the characteristics of the method, that is the sensitivity of the method to the parameters and data present in eq. (13). First of all, the left side of eq. (13) is a function of the following parameters: observational data (V, B-V, V-R, u); constants (c and p); the variable R. This function has a zero $R_0$ at a value which is determined numerically; it is important to study the behaviour of such function, with respect to R. Fig. 1 gives the plot of the above function for $\delta$Cep; it is a well-behaved function whose zero can safely be determined. The plot varies of course according to the star, but the general shape is similar.

Since the terms B and $c \cdot C$ in eq. (12) are additive terms, their effect is to move the whole curve by a vertical shift, thus changing the value of $R_0$. This explains the importance of the loops for both B, the area in the plane (V, B-V) and C, the area in the plane (B-V, V-R). However, the observational data enter in these quantities only globally, that is all data together combinate to give B (and C), and this means that errors on individual points have less influence on the final value of B and C, and hence of $R_0$. However, the evaluation of B and C is very dependent on the regularity of the curve: if the noise of the data is high, and the data have a large scatter around the fitted curve in the planes ($\phi$, V), ($\phi$, B-V), ($\phi$, V-R) then the area of B and C may be not well determined. Fig. 2 shows the plot of the magnitude-colour and colour-colour loops for $\delta$Cep. From this figure it is clear that the scatter in colour-colour loop is more critic with respect to that of the colour-magnitude loop, so that the term $\Delta B$ will be more influenced by the eventual poor quality of the data. In any case, since the area described by the colour-colour loop is very small (due to the characteristic of $\Delta B$ to be a correction term) it is reasonable to expect that the error introduced in this way will be very small; an occurrence already noticed by Sollazzo et al. (1981), (see their Fig. 8).

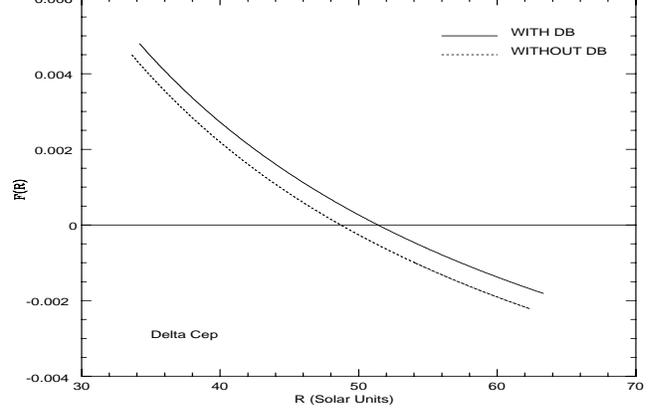

**Fig. 1.** Value of the solving function vs the radius of the star $\delta$ Cep

For what concerns the dependence on the converiosn factor p, Fig 3 shows the value of $R_0$ obtained for $\delta$Cep, W Sgr and SZ Tau, for five different values of p: a change of 0.05 in p corresponds to a change of 3.5% in $R_0$.

## 4. Application of the method

### 4.1. The samples

In order to apply the method, we searched the literature for hight quality sets of observational data. We considered mainly two large and homogeneous sets of data: one from Moffett & Barnes (1980, 1984) (MB hereafter, BVRI photometry), Barnes et al. (1987,1988) (BMS hereafter), Wilson et al. (1989) (WCBCM hereafter, radial velocities), and the other from Bersier et al. (1994) (BBB hereafter, Geneva System Photometry), Bersier et al. (1994) (BBMD hereafter, CORAVEL radial velocities).

There were 26 variables in common between the two sets; in these cases, in general, we used photometry from MB (we proceed in this way either because these data was generally more accurate than those in the Geneva system, or because the presence of the V-R colour allowed us to use our modified CORS method) and radial velocity data from BBMD (since these set of data were much more accurate than the other one). For other three stars: SY Cas, SY Nor and TW Nor for which are present radial velocity data from BBMD, but no photometry in the Geneva System, we used photometric data from Berdinikov (1992a,b,c, B in Tab. 1) for the first two, and Madore (1975, M in Tab. 1) for the third. The incompleteness of photometric data for SY Nor did not allow us to determine the radius of this star.

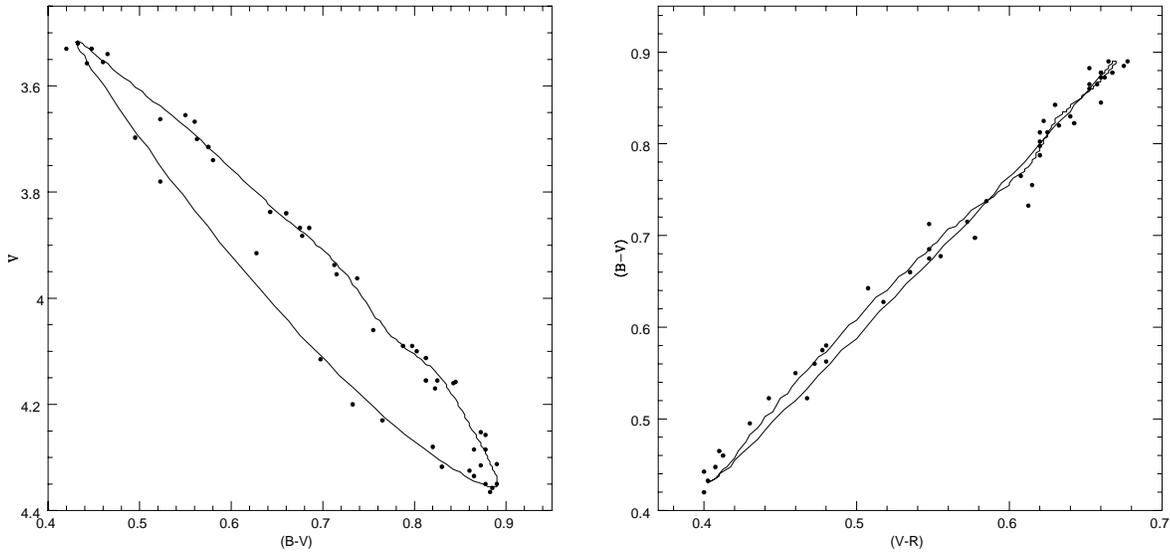

**Fig. 2.** Left: Loops in the plane (B-V, V) for $\delta$ Cep; Right: as in Left, but for the plane (V-R, B-V)

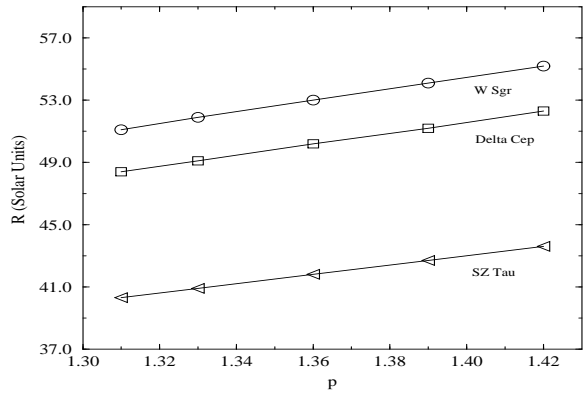

**Fig. 3.** Dependence of CORS output radii as a function of the conversion factor p, for three stars: Delta Cep, SZ Tau, T Vul

### 4.2. Particular stars

There were eight stars from BMS and WCBCM whose radial velocities curves were too poor to be used, they are: FF Aql, RU Cam, RW Cas, TU Cas, SU Cyg, AU Peg, VX Pup, S Sge.

For other two stars of the same sample: RW Cam and VY Cyg, the method did not reach the convergence, probably due to the poor quality of the data.

SW Tau was excluded because it is a Population II Cepheid.

From the BBMD sample we excluded from our computation V440 Per, which have convergence problems; CO Aur and V367 Sct because they are double or triple mode; the double stars DL Cas and V465 Mon because of problems in separating the orbital motion from the radial velocity curve.

### 4.3. Double stars

From the literature is possible to find indications about the doubleness of several cepheids; in particular, following BBB, BBMD and Pont et al. (1994) the following variables result to be double: SU Cas, DD Cas, VZ Cyg, DX Gem, $\zeta$Gem, RZ Gem, Z Lac, T Mon, S Nor, SV Per, Y Sct, U Sgr, W Sgr. Among these double stars, only U Sgr shows an anomaluos position in Fig 4 (see below).

## 5. Results

In Table 1 we report the results of our analysis for all stars of our sample for which the CORS method reached the convergence; from left we report in the order: name of the cepheid; period; radial velocity data source; photometric method (Gieren et al. 1989); radius obtained with CORS (without $\Delta B$ term); as before but with the $\Delta B$ term.

**Table 1.** Results from our method and comparison with previous determinations with the Surface Brightness Method by Gieren et al. (1989). The last two columns show the radius determinated in this work.

| Cepheid | Period (days) | Photometric source | Rad. Vel. source | Surf. Brigh. radius (R$_\odot$) | CORS radius (R$_\odot$) | new CORS radius (R$_\odot$) |
|---|---|---|---|---|---|---|
| U Aql | 7.024100 | MB | BMS | 54.65 | 55.5 | 61.3 |
| $\eta$ Aql | 7.176779 | MB | BMS | 54.93 | 57.1 | 56.7 |
| FM Aql | 6.114240 | MB | BMS | 54.82 | 58.8 | 60.6 |
| TT Aql | 13.755290 | MB | BMS | 97.62 | 86.0 | 95.3 |
| V496 Aql | 6.807164 | MB | BMS | 45.52 | 39.2 | 36.4 |
| RT Aur | 3.728220 | MB | BMS | 32.90 | 41.6 | 42.2 |
| SY Aur | 10.144698 | MB | BMS | 61.59 | 69.4 | 66.3 |
| RX Cam | 7.912190 | MB | BMS | 76.0 | 57.3 | 58.9 |
| CF Cas | 4.87514 | MB | BBMD |  | 46.9 | 50.2 |
| DD Cas | 9.81274 | MB | BBMD | 85.12 | 72.5 | 79.2 |
| FM Cas | 5.809232 | MB | BBMD | 62.05 | 49.9 | 49.9 |
| SU Cas | 1.949317 | MB | BBMD |  | 29.2 | 32.5 |
| SY Cas | 4.07110 | B | BBMD |  | 48.0 | 52.7 |
| V636 Cas | 8.375490 | BBB | BBMD |  | 74.1 |  |
| $\delta$ Cep | 5.366269 | MB | BBMD | 41.60 | 50.2 | 52.8 |
| CR Cep | 6.232870 | MB | BBMD |  | 58.6 | 56.9 |
| X Cyg | 16.385692 | MB | BBMD | 118.13 | 89.6 | 96.2 |
| DT Cyg | 2.499086 | MB | BBMD |  | 37.9 | 42.9 |
| SZ Cyg | 15.109642 | MB | BMS | 117.74 | 83.8 | 87.3 |
| VZ Cyg | 4.8644 | MB | BBMD | 37.17 | 44.4 | 47.3 |
| V386 Cyg | 5.257655 | MB | BMS | 41.86 | 38.1 | 40.8 |
| V532 Cyg | 3.283612 | MB | BMS |  | 38.2 | 45.2 |
| RZ CMa | 4.254926 | MB | BMS |  | 39.4 | 42.8 |
| RY CMa | 4.678425 | MB | BMS | 43.11 | 33.1 | 35.8 |
| TW CMa | 6.995374 | MB | BMS | 57.28 | 63.3 | 64.8 |
| $\zeta$ Gem | 10.149955 | MB | BBMD | 64.94 | 73.5 | 86.2 |
| W Gem | 7.913960 | MB | BMS | 50.66 | 56.6 | 60.7 |
| BB Gem | 2.308207 | BBB | BBMD |  | 31.4 |  |
| DX Gem | 3.136379 | MB | BBMD |  | 41.8 | 41.1 |
| RZ Gem | 5.529162 | MB | BMS | 52.49 | 61.4 | 68.2 |
| X Lac | 5.44487 | MB | BBMD | 64.71 | 74.3 | 71.2 |
| Y Lac | 4.323776 | MB | BMS | 50.27 | 50.7 | 51.7 |
| Z Lac | 10.8866000 | MB | BMS | 68.88 | 83.8 | 86.8 |
| BG Lac | 5.331938 | MB | BMS | 45.42 | 36.9 | 35.8 |
| RR Lac | 6.416190 | MB | BBMD | 45.61 | 45.1 | 46.4 |
| BE Mon | 2.70551 | BBB | BBMD |  | 29.7 |  |
| T Mon | 27.022600 | MB | BBMD | 172.19 | 170.5 | 188.3 |
| CV Mon | 5.378793 | MB | BMS |  | 43.1 | 49.2 |
| SV Mon | 15.232780 | MB | BMS | 100.64 | 119.6 | 130.7 |
| V508 Mon | 4.133608 | BBB | BBBMD |  | 56.1 |  |
| S Nor | 9.764244 | BBB | BBMD |  | 57.2 |  |
| TW Nor | 10.78531 | M | BBMD |  | 67.5 |  |
| V340 Nor | 11.28871 | BBB | BBMD |  | 71.6 |  |
| Y Oph | 17.126780 | MB | BMS | 71.88 | 123.3 | 112.2 |
| BF Oph | 4.067695 | MB | BMS | 35.87 | 31.2 | 32.4 |
| GQ Ori | 8.616127 | MB | BMS | 72.44 | 54.6 | 54.7 |
| AW Per | 6.463720 | MB | BMS | 47.34 | 54.6 | 54.8 |
| SV Per | 11.129318 | MB | BMS | 63.97 | 65.9 | 79.9 |
| VX Per | 10.889040 | MB | BMS | 77.27 | 63.7 | 62.1 |
| X Pup | 25.961000 | MB | BMS | 118.04 | 156.1 | 164.1 |
| RS Pup | 41.415000 | MB | BMS | 262.60 | 205.7 | 207.9 |
| AQ Pup | 29.839810 | MB | BMS | 197.20 | 124.8 | 145.9 |
| WX Pup | 8.937050 | MB | BMS | 76.54 | 66.1 | 63.1 |
| RV Sco | 6.061388 | MB | BMS | 49.29 | 42.2 | 48.3 |
| V500 Sco | 9.316862 | MB | BMS | 53.12 | 63.3 | 72.8 |
| Y Sct | 10.341650 | MB | BMS | 83.52 | 62.4 | 71.5 |
| EV Sct | 3.091047 | MB | BBMD |  | 30.9 | 36.6 |
| U Sgr | 6.745363 | BBB | BBMD | 60.03 | 34.5 |  |
| W Sgr | 7.594935 | MB | BBMD | 63.28 | 53.0 | 58.6 |
| X Sgr | 7.012630 | MB | BMS | 49.77 | 56.5 | 58.6 |

| Cepheid | Period (days) | Photometric source | Rad. Vel. source | Surf. Brigh. radius (R☉) | CORS radius (R☉) | new CORS radius (R☉) |
|---|---|---|---|---|---|---|
| Y Sgr | 5.773400 | MB | BMS | 50.04 | 60.7 | 61.4 |
| BB Sgr | 6.637117 | MB | MB | 42.12 | 50.0 | 53.5 |
| AP Sgr | 5.057936 | MB | BMS | 44.02 | 42.6 | 50.0 |
| YZ Sgr | 9.553606 | MB | MB | | 84.2 | 136.0 |
| V350 Sgr | 5.154557 | MB | BMS | 47.80 | 41.8 | 46.0 |
| EU Tau | 2.102182 | BBB | BBMD | | 20.6 | |
| ST Tau | 4.034299 | MB | BMS+BBMD | 41.41 | 46.6 | 47.1 |
| SZ Tau | 3.149138 | MB | BBMD | 37.83 | 41.8 | 44.8 |
| T Vul | 4.435453 | MB | BBMD | 38.24 | 48.8 | 48.8 |
| U vul | 7.990821 | MB | BBMD | 56.54 | 57.9 | 61.1 |
| X vul | 6.31949 | MB | BBMD | 46.24 | 61.5 | 68.4 |
| SV Vul | 45.00061 | BBB | BBMD | 202.15 | 189.5 | |

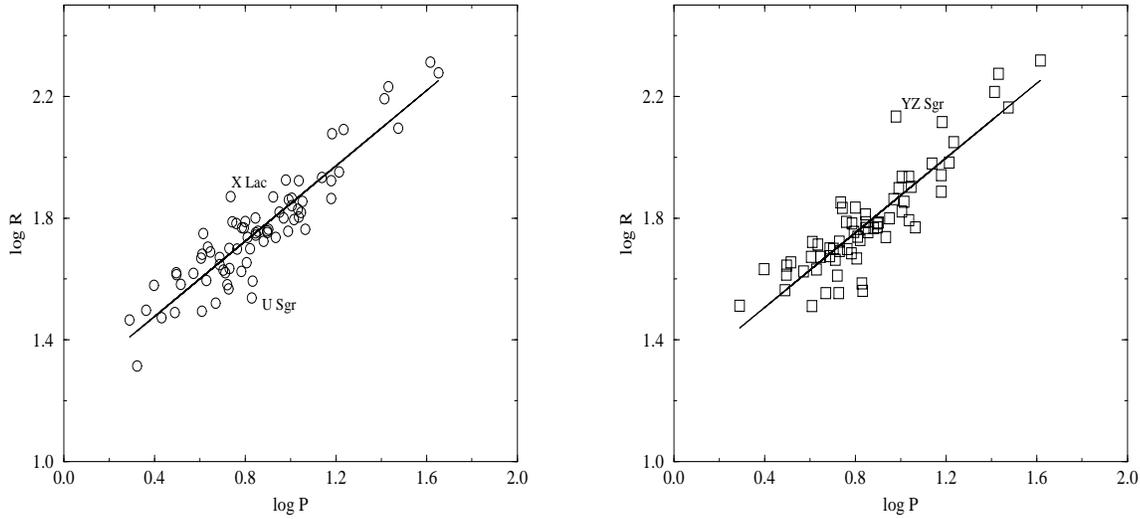

**Fig. 4.** Left: Period-Radius relation obtained from CORS method without $\Delta B$; Right: as before, but with $\Delta B$.

**Table 2.** Comparison of coefficients of Period-Radius relation ($\log R = a \log P + b$) between this paper and selected other works available in literature

| Method | a | b | Source |
|---|---|---|---|
| Theory | 0.692 ± 0.006 | 1.179 ± 0.006 | Fernie(1984) |
| Theory | 0.70 | 1.17 | Cogan (1978) |
| Theory | 0.72 | 1.07 | Karp (1975) |
| Surf. Bright. | 0.743 ± 0.023 | 1.108 ± 0.023 | Gieren et al. (1989) |
| Surf. Bright. | 0.751 ± 0.026 | 1.070 ± 0.008 | Laney & Stobie (1995) |
| LMC cepheids | 0.716 ± 0.010 | 1.139 ± 0.009 | Di Benedetto (1994) |
| CORS without $\Delta B$ | 0.622 ± 0.029 | 1.226 ± 0.026 | This paper |
| CORS with $\Delta B$ | 0.606 ± 0.037 | 1.263 ± 0.033 | This paper |

These data are plotted in Fig 4 were we present the Period-Radius relation as a result of CORS method in two cases, with and without $\Delta B$ respectively; the solid lines superimposed represent a least square fit to our data that leeds to the following Period-Radius relation (74 cepheids) without the $\Delta B$ term in eq. (13):

$$\log R = (0.619 \pm 0.032) \log P + (1.229 \pm 0.028) \quad (14)$$

If we exclude from our fit X Lac (it shows a clear phase shift) and U Sgr (double star) we obtain a slightly better error:

$$\log R = (0.622 \pm 0.029) \log P + (1.226 \pm 0.026) \quad (15)$$

Now, if we consider the $\Delta B$ term in eq. (13) for cepheids with BVRI photometric observations (65 stars) we obtain:

$$\log R = (0.614 \pm 0.040) \log P + (1.260 \pm 0.036) \quad (16)$$

And excluding the large scattering YZ Sgr (poor radial velocity data) we obtain:

$$\log R = (0.606 \pm 0.037) \log P + (1.263 \pm 0.033) \quad (17)$$

Our Period-Colour relations are reported also in Tab. 2 in comparison with other selected results from the literature. We note that our P-R relations show a slope slightly shallower than either the more recent determinations via Baade-Wesselink methods, or theoretical determination. Moreover the use of a second colour (i.e. the inclusion of $\Delta B$ in the determination of radius) goes in the sense to reduce further on the slope. Since we use (at least partially) the same data of Gieren et al. (1989) it is surprising to find such a different result. Anyway, in their recent paper, Laney and Stobie (1995) applied a Surface Brightness method to a set of 31 Cepheids, using different pairs of magnitudes and colours, from (V,B-V) to (K,J-K), and found a dependence of the P-R relation from the colours used, in the sense that the slope is shallower if "blue" colours are used instead of the infrared ones. This could partially explain our shallower slope in the P-R relations, since our two determinations of R depend either from two or three "blue" colours.

## 6. Conclusions

We have presented a method for computing Cepheid radii, which merges the CORS and the Surface Brightness method into a single, easy to use method applicable to observations in different photometric systems.
An interesting feature of this method is its indipendence from the konwledge of the reddening corrections, at least within the limits of an assumpion that the derivative of $(V-R)_0$ is the same as the one of (V-R).

hod is efficient and gives coherent results; we were able to compute radii for 74 Cepheids with a rather straightforward procedure. However, the improvements in the P-R relation are almost negligible with respect to previous works at least in terms of the scatter of the data. This scatter can be explained by the fact that our $\Delta B$ is only an approximation to the correct $\Delta B$ of the CORS method; however, its presence gives the right trend to the slope of the P-R relation, which is lower with $\Delta B$ than without $\Delta B$, as found also with the correct $\Delta B$ (see Sollazzo et al. 1981). The fact that we always use three bands (B,V,R in this paper) should also give more consistency to the results.

Further work is needed to improve the approximation, but the basis of the method seems now well laid. A possible idea, wich we will work on, is in computing $\Delta B$ as a sum of two terms, one based on the observations (like in the present paper) and one based on theoretical computations from Kurucz's models (like the original CORS method did, but only for a particular photometric system). We are also working on a method to compute statistically meaningful errors on the radii, based on the quality of the observations, rather than on the purely numerical errors of the fitting procedure.

*Acknowledgements.* We would like to thank Prof. T.G. Barnes for having provided us with his data in computer form. We thank also Dr. Giuseppe Bono for a critical reading of the manuscript.

## References

Barnes, T.G., Evans, D.S. 1976, MNRAS, 174, 489
Barnes, T.G., Moffett, T.J., Slovak, M.H. 1987, ApJS, 65, 307
Barnes, T.G., Moffett, T.J., Slovak, M.H. 1988, ApJS, 66, 43
Berdinikov, L.N., Turner, D.G. 1995, Pis'ma Astron. J., in press
Berdinikov, L.N. 1992, A&A Transactions, 2, 1
Berdinikov, L.N. 1992, A&A Transactions, 2, 47
Berdinikov, L.N. 1992, A&A Transactions, 2, 107
Bersier, D., Burki, G., Burnet, M. 1994, A&AS, 108,9
Bersier, D., Burki, G., Mayor, M. Duquennoy, A. 1994, A&AS, 108, 25
Caccin, B., Onnembo, A., Russo, G., Sollazzo, C. 1981, A&A, 97, 104
Di Benedetto, G.P. 1994, A&A, 285, 819
Fernie, J.D. 1984, ApJ, 282, 641
Gautschy, A. 1987, Vistas Astron., 30, 197
Gieren, W.P. 1986, MNRAS, 222, 251
Gieren, W.P., Barnes, T.G., Moffett, T.J. 1989, ApJ, 342, 467
Karp, A.H. 1975, ApJ, 199, 448
Laney, C.D., Stobie, R.S. 1995, MNRAS, 274, 337
Madore, B.F. 1975, ApJS, 29, 219
Moffett, T.J., Barnes, T.G. 1980, ApJS, 44, 427
Moffett, T.J., Barnes, T.G. 1984, ApJS, 55, 389
Moffett, T.J., Barnes, T.G. 1987, ApJ, 323, 280
Onnembo, A., Buonaura, B., Caccin, B., Russo, G., Sollazzo, C. 1981, A&A, 97, 104

Pel, J.W. 1976, A&AS, 24, 413
Pel, J.W. 1978, A&AS, 62, 75
Pont, F., Mayor, M., Burki, G. 1994, A&A, 285, 415
Russo, G., Sollazzo, C., Coppola, M. 1981, A&A, 102, 20
Sollazzo, C., Russo, G., Onnembo, A., Caccin, B. 1981, A&A, 99, 66
Wesselink, A.J. 1946, BAN, 368, 91
Wilson, D.W., Carter, M.W., Barnes, T.G., Van Citters, W.G., Moffett, T.J. 1989, ApJS, 69, 951